\documentclass[fleqn,usenatbib]{mnras}

\usepackage[T1]{fontenc}
\usepackage{graphicx}	
\pdfoutput=1
\usepackage{amsmath}	
\usepackage{amssymb}	
\usepackage{natbib} 
\usepackage[english]{babel}
\usepackage{lineno,hyperref}
\usepackage{float}
\usepackage{xcolor}
\usepackage{booktabs}
\usepackage{listings}
\usepackage{algpseudocode}
\usepackage[linesnumbered,ruled,vlined]{algorithm2e}
\usepackage[normalem]{ulem}
\usepackage{graphicx, subcaption, caption}

\definecolor{codegreen}{rgb}{0,0.6,0}
\definecolor{codegray}{rgb}{0.5,0.5,0.5}
\definecolor{codepurple}{rgb}{0.58,0,0.82}
\definecolor{backcolour}{rgb}{0.95,0.95,0.92}
 
\lstdefinestyle{mystyle}{
    backgroundcolor=\color{backcolour},   
    commentstyle=\color{codegreen},
    keywordstyle=\color{magenta},
    numberstyle=\tiny\color{codegray},
    stringstyle=\color{codepurple},
    basicstyle=\ttfamily\footnotesize,
    breakatwhitespace=false,         
    breaklines=true,                 
    captionpos=b,                    
    keepspaces=true,                 
    numbers=left,                    
    numbersep=5pt,                  
    showspaces=false,                
    showstringspaces=false,
    showtabs=false,                  
    tabsize=2
}
 
\DeclareRobustCommand{\VAN}[3]{#2}
\let\VANthebibliography\thebibliography
\def\thebibliography{\DeclareRobustCommand{\VAN}[3]{##3}\VANthebibliography}

\title[RFI characterisation]{Multidimensional RFI Framework for Characterising Radio Astronomy Observatories}

\author[I. Sihlangu et al.]{
Isaac Sihlangu$^{1,2}$, Nadeem Oozeer$^{1,3}$
and  Bruce  A. Bassett$^{1,2,3,4}$ 
\newauthor 
\\~\\
$^{1}$  South African Radio Astronomy Observatory, 2 Fir Street, Observatory, Cape Town, 7925, South Africa\\
$^2$ Department of Maths and Applied Maths, University of Cape Town, Rondebosch, Cape Town,7700, South Africa\\
$^{3}$ African Institute for Mathematical Sciences, 6 Melrose Road, Muizenberg, 7945, South Africa\\
$^{4}$ South African Astronomical Observatory, Observatory, Cape Town, 7925, South Africa\\ 
}

\date{Accepted XXX. Received YYY; in original form ZZZ}

\begin{document}
\label{firstpage}
\pagerange{\pageref{firstpage}--\pageref{lastpage}}
\maketitle

\begin{abstract}
Radio Frequency Interference (RFI) has historically plagued radio astronomy, worsening with the rapid spread of electronics and increasing telescope sensitivity. 
We present a multi-dimensional probabilistic framework for characterising the RFI environment around a radio astronomy site that uses  automatically flagged data from the array itself. We illustrate the framework using about 1500 hours of commissioning data from the MeerKAT radio telescope; producing a 6-dimensional array that yields both average RFI occupancy as well as confidence intervals around the mean as a function of key variables (frequency, direction, baseline, time). Our results provide the first detailed view of the MeerKAT RFI environment at high sensitivity as a function of direction, frequency, time of day and baseline. They allow us to track the historical evolution of the RFI and to quantify fluctuations which can be used for alerting on new RFI. As expected we find the major RFI contributors for MeerKAT site are from Global Positioning System (GPS) satellites, flight Distance Measurement Equipment (DME) and the Global System for Mobile (GSM) Communications. Beyond characterising RFI environments our approach allows observers access to the prior probability of RFI in any combination of tracked variables, allowing for more efficient observation planning and data excision.
\end{abstract}




\section{Introduction}

Radio signals from astronomical sources are extremely weak and are easily corrupted or overwhelmed by man-made radio signals such as cellphones, satellites, aircraft and telescope electronics. Any radio signal other than the desired astronomical signal is called an unwanted signal, or spurious radiation and is classified as Radio Frequency Interference (RFI). RFI is increasingly threatening radio observatories due to our increasingly technological society, \cite{fridman2001rfi}.\par\bigskip

The MeerKAT radio telescope, referred to as MeerKAT onward, is amongst the most sensitive L-band radio telescope of its kind and is observing the radio sky with unprecedented depth and detail,  \cite{camilo2018revival}. Building such an instrument comes at the cost of also picking up very faint RFI sources. The MeerKAT L-band frequency range is dominated with known RFI sources from Global System for Mobile (GSM) communications, flight Distance Measurement Equipment (DME) from aircraft and Global Position System (GPS) satellites.\par\bigskip

Although radio astronomy has been carried out for decades, we have not seen much framework that collects and characterises the RFI environment from the telescope observation measurements. If available, such frameworks are internal to the observatory and are rarely accessible. Radio astronomers generally flag the outliers from their science data without caring much about their causes. These RFI flags are discarded and rarely fed back to the observatory. We therefore propose a framework that can allow any radio astronomy site to perform the task of keeping track of these RFI from the huge amount of data that has been collected.\par\bigskip

Our proposed framework, investigates the RFI occupancy surrounding the MeerKAT site using a probabilistic approach. For each observation file we produce the probability of RFI as function of various parameters. The paper is divided as follows. Section 2 describes the MeerKAT in-house RFI detection methodologies. Section 3 gives the framework used for analysing of the RFI occupancy, including the algorithm design and statistical methods. Section 4 provides the results and discussions followed by a conclusion in Section 5.

\section{MeerKAT SDP RFI flagger}
In order to understand our new proposed framework, we begin by giving an overview of the MeerKAT Science Data Processing (SDP\footnote{SDP - Science Data Processing: Is the MeerKAT team which is responsible for quality control and quality assurance of the data.}) pipeline. The MeerKAT receiver captures the radio signal and converts it into voltages which is then filtered and amplified, (\cite{asad2019primary}). The amplified voltage is then digitised (at the receptor) and sent to the correlator/beamformer (CBF) situated at the Karoo Array Processor Building (KAPB) via underground optical fibres (\cite{asad2019primary}), that is located 12 km away from the MeerKAT core. The correlator implements the FX/B signal processing,  \cite{camilo2018revival}. As explained in \cite{mauch20201}, the F-engine coarsely aligns the voltages and corrects for both geometrical and instrumental delays and splits the data into frequency channels. The aligned voltages from pairs of antennas can then undergo various processes such as the correlation of the signal by the X-engine or beamforming of the signal by the B-engine,~\cite{mauch20201}. The raw visibilities are further piped into the ingest at 0.5s dump period which then processes the data to produce the  visibility data product (which is usually averaged to 8s or according to the user's observing parameters - known as $L0$). A calibration pipeline is run on the $L0$ visibilities to produce the calibrated visibility that is called the  $L1$ data product.\par\bigskip

RFI detection in the MeerKAT SDP pipeline happens at two stages, that is at the ingest step and at the calibration. The high time resolution RFI detection happens during the ingest step. Here, the strong RFI is detected by checking for outliers in the frequency axis in individual correlator dumps. At this stage averaging of data is carried out. The samples which are detected as RFI are excised and only unflagged samples are averaged in time as per the observation requirement and further used in the data processing pipeline. The output of the ingest step is, therefore, an averaged RFI excised data-set with pertinent meta-data stored in $telestate$\footnote{$Telestate$ is a redis database from telescope state (telstate) that contains meta-data.}.\par\bigskip

To account for data loss due to the ingest excision, each visibility data point has an associated weight ($W_{SDP}$) that allows us to calculate how many samples were averaged to produce the visibility. If we define $N$ as number of correlator samples, $V_{CBF}$ as the visibility sample from the correlator/beamformer (CBF), $U$ as the set of indices of unflagged correlator visibilities, we can then calculate the SDP visibility sample $V_{SDP}$ using the following:\par\bigskip

\begin{eqnarray}
V_{SDP} = W_{SDP} \sum_{i \in U}^N V_{CBF[i]}
\end{eqnarray}

where $i$ represents correlator/beamformer sample  index  and $W_{SDP} =  \frac{1}{N_{U}}$, with $N_U$ being defined as the number of unflagged samples by the ingest. The ingest flags become \textbf{True} when all $N$ samples are flagged as RFI by the ingest, then at this point there is no excision of data. With partial flagging or excision of data, the ingest flags becomes \textbf{False}. The ingest RFI detection algorithm usually only detect narrow regions around the brightest RFI spikes in the data, and further flagging is required.\par\bigskip

On the other hand, during the calibration step, the MeerKAT in-house developed RFI flagger\footnote{https://github.com/ska-sa/katsdpsigproc} (hereafter called the SDP flagger) is used. The SDP flagger is based on the variation of VarThreshold method used in the classic  $AOFlagger$ algorithm,~\cite{Offringa2010post}. The SDP flagger works on a quasi-real-time model. It runs on a two-dimensional data array of time and frequency. As already mentioned in the introduction, the MeerKAT L-band frequency is mostly corrupted by RFI from the GSM, DME and GPS sources. Therefore, a static mask was developed to flag at all times the data for those frequency ranges. This mask is applied only on short baselines ($\leq$1000m), because of the RFI dependency on baseline length. We will discuss this in details later in this paper. The unmasked visbility data is loaded into the SDP flagger per scan, where a scan is defined as a collection of SDP visibilities over a certain time period. For MeerKAT, a scan is on average between 5 - 15 minutes. Furthermore, the algorithm treats each baseline\footnote{A baseline is the vector joining any 2 antenna pairs. For N antennas, the total baseline is $N(N-1)/2$.} in each scan independently thus allowing the parallelisation of the algorithm along the baseline axis.\par\bigskip

A smooth background fit is applied to the unmasked data by convolving it with a 2-D Gaussian whose widths are larger than expected RFI spike widths in both time and frequency and are smaller than any variations in the bandpass or changes in amplitude with time,~\cite{mauch20201}. This ensures that the smooth background ignores any potential RFI spikes in the data but follows the true shape of the background. Data already flagged from the ingest or from the static RFI mask are given zero weight and therefore do not contribute to the background estimation.\par\bigskip

The fitted smooth background is subsequently subtracted from the data, and the standard deviation is measured from the masked residual. This standard deviation is used as the basis for the threshold for spike detection. First, the data are averaged in time over the whole scan and outliers in the resulting 1-D frequency spectrum are located; this is to find faint spikes in the time axis that would otherwise be missed. RFI channels found in the 1-D spectrum are flagged for all times in the scan. Finally, the full data are flagged in the time and frequency dimensions.\par\bigskip

\section{KATHPRFI Framework}

As mentioned previously, RFI is a nuisance for astronomers as they corrupt the relatively weak radio signal. These unwanted signals are normally removed/flagged by masking out the regions around what the astronomer decides as RFI. Such excision can be very subjective. Currently there is no way to keep track of RFI or to quantify the RFI health for the environment around the MeerKAT/SKA site. During the commissioning and testing phases, huge amount of data are available from MeerKAT. We therefore carried out a statistical analysis of the RFI environment as measured by the MeerKAT telescope using the  Karoo Array Telescope Historical Probability Radio Frequency Interference (KATHPRFI) framework.\par\bigskip

One of the the goals of the KATHPRFI framework is to provide  MeerKAT users with a tool that will aid them to keep track of changes in the RFI statistics over a long period as measured by the telescope. Such information is very useful for various users such as astronomers, telescope operators, RFI Engineers or anyone interested in the RFI health of the observatory. For astronomers, having a better understanding of the RFI environment is vital  in the preparation of observation proposals and also to carry out scientific analysis of their experiment. On the other hand, the RFI statistics can also allow us to build an intelligent observation scheduler and monitoring system for the RFI on site. The proposed tool can thus help the telescope operation team to understand the RFI environment on site and also aid them to detect system failures and any other telescope electronics issues.

\subsection{KATHPRFI - Design}

We chose an evolutionary prototyping model in our design instead of a throw-away approach. The evolutionary prototyping is a life cycle model wherein the concept of the system is developed as the project progresses,~\cite{carter2001evolving}. The evolutionary prototyping approach allows easy modification of the system in response to the user’s inputs. Our motivation behind choosing an evolutionary prototyping approach instead of the throw-away approach is due to the complicated nature of RFI signals, as it is difficult to frame the specifications of our system from the word go.\par\bigskip

The KATHPRFI framework retrieves the visibility data from the archive, followed by running the SDP flagger on the visibility data set. The SDP flagger was run offline to produce the required RFI flag files since the online SDP flagger  produces flags with a static mask. The static mask is the masking of some part of the MeerKAT bandpass due to the known RFI transmitters that have 100$\%$ duty cycle and band usage. Thus, the static mask is not determined by the data, hence its name. Therefore, running the SDP flagger offline allowed us to remove the static mask so that we can get the real RFI detected flags without the mask.\par

Using the offline flags the KATHPRFI framework then construct a \textbf{\textit{Master}} and a \textbf{\textit{Counter}} array, which contains the descriptive statistics about the RFI from each visbility dataset. The \textbf{\textit{Master}} array contains the number of RFI points per voxel, whereas the \textbf{\textit{Counter}} array contains the total number of observations per voxel. Both arrays are built around the concept of a 6-D array. The first dimension is the filename, from which we can extract the date of the observation. The other 5 dimensions are; time of the day (T), frequency (F), baseline length (B), elevation (El) and azimuth (Az). The shape of the 5-D data array is [24$\times$4096$\times$2016$\times$8$\times$24].

\subsection{KATHRFI - Algorithm}

\begin{figure*}
\centering
    \includegraphics[width=17cm]{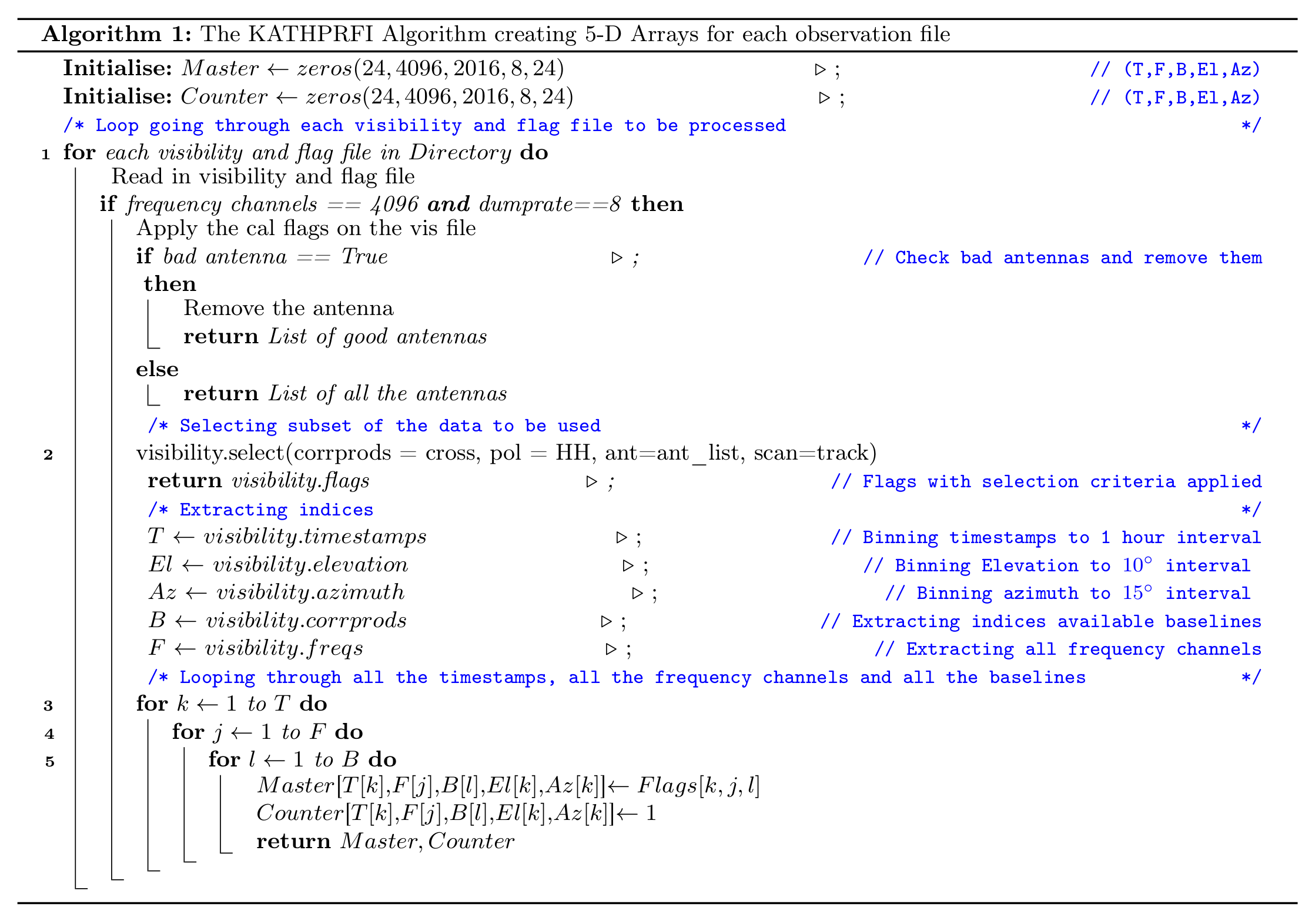}
    \caption{The KATHPRFI Algorithm creating 5-D Arrays for each observation file.}
\label{alg0}
\end{figure*}

The KATHRFI script starts by initialising the \textbf{\textit{Master}} and the \textbf{\textit{Counter}} arrays, as depicted by the algorithm in Fig \ref{alg0}. Block 1 of the algorithm reads in the visibility file and the offline flag file, followed by applying the offline flags. The final step of Block 1 is the pre-processing stage whereby we remove any bad antennas from the data.  An antenna is flagged as a bad if, during an observation, it fails for some reasons. The $katdal$ \footnote{\url{https://github.com/ska-sa/katdal}} library is used to get the information about the antenna activities during the observation. If we find a STOP state on the activity lists of an antenna during an observing track, we would flag that specific antenna as a bad and remove it from the analysis.\par\bigskip

In the second block, we choose the parameters of interest to produce a subset of the data. The $katdal$ library allows us to do the selection. Thereafter, the flag array is returned with the applied selection criteria.\par\bigskip

Due to computational limitation, the  data are binned. This is carried out in the final step of Block 2. This causes the full resolution of certain attributes, such as the time of observation, azimuth and elevation, not to be stored. We have managed to maintain the full resolution of the frequency and the baseline axis. Time is hence binned per hour into 24 hours of a day. The elevation is binned into $8^\circ$ intervals and the azimuth is binned in $15^\circ$ intervals. Any mention of Azimuth, Elevation and Time hereafter will mean the respective bin value and not the actual value.\par\bigskip

Block 3 to Block 5 are nested loops that go over timestamps, frequency channels and baselines respectively. This is done to update the \textbf{\textit{Master}} and the \textbf{\textit{Counter}} array based on the indices extracted from a particular observation file in question. If  for example an observation started at 10:07 and ran until 10:11, the KATHRFI script will put all of the data for that time in the $10^{th}$ hour bin. It will also check which antennas were present during the observation and update the baseline array accordingly.

\subsection{Statistical methods to calculate the averages}\label{analysis}

Randomly chosen imaging observations were used to create the data set, that we will call Historical probability Data Release one (HPDR1). The observation dataset used is equivalent to 1500 hours of observing time ($\sim$ 200 TB)  which was collected from May 2018 to December 2018. These observations were carried out in the L-band, containing 4096 channels with a 208.984 kHz channel width. As mentioned previously, the datasets are run through the in-house RFI detection algorithm and the respective flag tables were created.\par\bigskip

In order to calculate the probabilities, we adopt the following approach. Suppose that $\alpha$ is the number of RFI samples as obtained from the \textbf{\textit{Master}} array and $\beta$ is the number of NON-RFI samples (i.e. Total of \textbf{\textit{Counter}} array - Total of \textbf{\textit{Master}} array); where  \textbf{\textit{Counter}} array is the total number of observed samples. Then we can compute the probability estimate, P(RFI), in a voxel as follows: 

\begin{eqnarray}
P(RFI|T,F,B,El,Az) = \frac{\alpha_{T,F,B,El,Az}}{\alpha_{T,F,B,El,Az}+\beta_{T,F,B,El,Az}}
\end{eqnarray}

\noindent
where $T, F, B, El$ and $Az$ are the indices of time of the day, frequency, baseline length, elevation and azimuth in a given voxel respectively. In order for us to compute the probability of RFI for a given dimension, we need to marginalise over all other dimensions. For instance, if we want to compute the probability of observing RFI as a function of the frequency, we sum both \textbf{\textit{Master}} and \textbf{\textit{Counter}} array in all other axes except the frequency axis, and then we divide one by the other, and the resulting array will be the probability of observing RFI as a function of frequency. Mathematically it can be written as,

\begin{eqnarray}\label{average}
P(RFI|F) = \frac{\sum_{T,B,El,Az}(\alpha_{F})}{\sum_{T,B,El,Az}(\alpha_{F}+\beta_{F})}.
\end{eqnarray}
\noindent
In order to calculate the average, we used two different methods. 
\begin{itemize}
    \item The first method was to update the  \textbf{\textit{Master}} and \textbf{\textit{Counter}} array every time we get a new observation. At the end using Equation \ref{average} we computed the average RFI probability, effectively combining all observations into a single long observation. This is referred to as the Combine Average (CA).
    \item The other method consists of computing the average of each individual file using Equation \ref{average} and finally computing the average of those probabilities. We call this method the Average of Average (AoA). 
\end{itemize}

These two averages will coincide if all the files have the same length or if the average of each file is the same, but in general the CA and AoA will differ.
\section{Results and discussion}
\begin{figure*}
\centering
    \includegraphics[width=12cm]{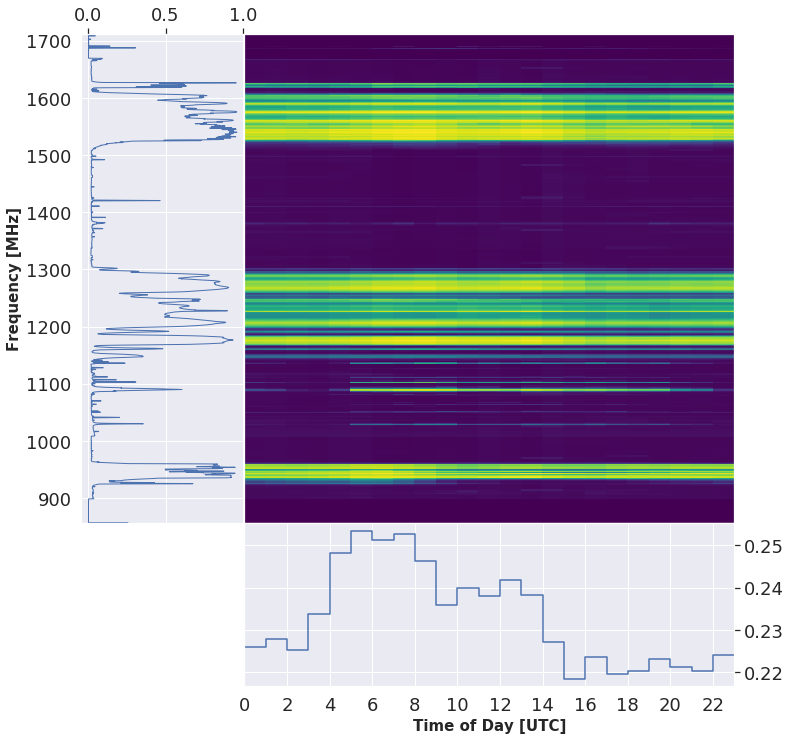}
    \caption{The distribution of RFI probability/occupancy as a function of frequency and time of the day. The average RFI over all frequencies and times of day is about 23$\%$. The colour scale indicates the amount of RFI detected in a specific time-frequency bin with yellow being the highest probability and purple the lowest probability.}
\label{fig:time_freq}
\end{figure*}

The overall RFI probability distribution picked up by MeerKAT in the HH polarization as a function of time of the day in Coordinated Universal Time (UTC) and frequency in megahertz (MHz) is shown in Fig. \ref{fig:time_freq}. Our results show a clear pattern between the hour of the day and the RFI probability. We see a drop of RFI probability during the night time (i.e. 18:00 - 04:00 UTC) as compared to the day time (i.e. 05:00 - 17:00 UTC).  A maximum variation of 4$\%$ is observed between hours of the day in the RFI occupancy with an average of 23$\%$. These results have confirmed that during the day time, the RFI probability is high as compared to the night time. Our analysis therefore allowed us to validate some of the claims and hypothesis using MeerKAT commissioning imaging observations.\par\bigskip

 \begin{figure*}
 \begin{subfigure}[t]{0.4\hsize}
    \includegraphics[width=1.0\linewidth]{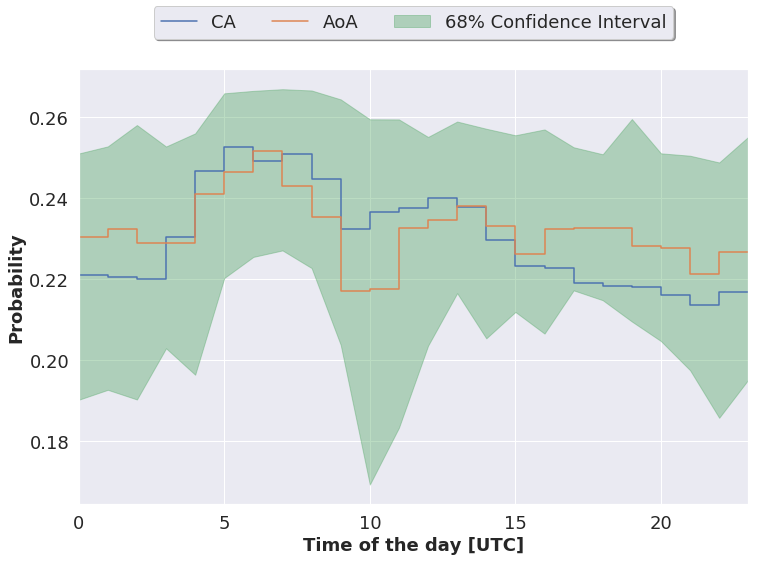}
    \caption{ The average RFI occupancy as a function of time of day.}
    \label{time}
\end{subfigure}
\begin{subfigure}[t]{0.4\hsize}
    \includegraphics[width=1.0\linewidth]{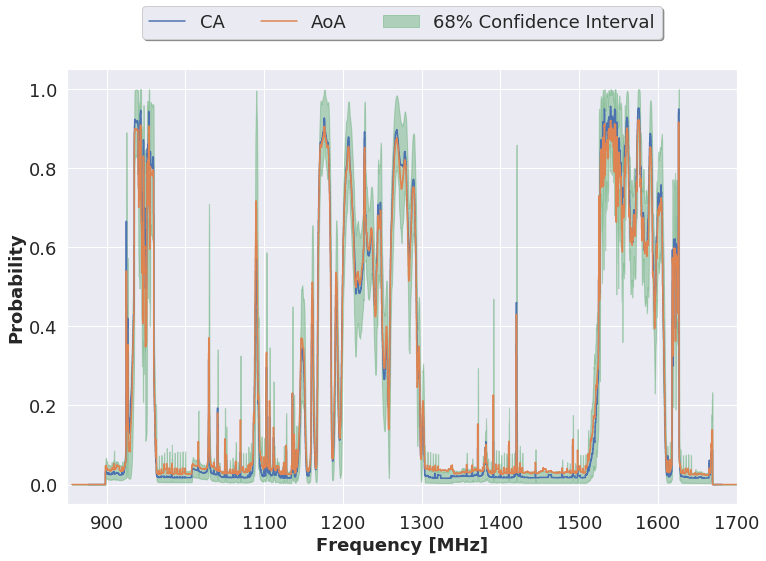}
     \caption{The distribution of RFI probability as a function of frequency.}
    \label{fig:freq_95_68}
\end{subfigure}   
    \caption{MeerKAT RFI occupancy as function of time of the day and frequency. The green region represents the 68$\%$ confidence interval computed over the historial observations. The blue line is the Combined Average (CA) and the orange line is the Average of Average (AoA) discussed in the text.}
\label{freq_time_rfi_dist}
\end{figure*}

We noticed that at 05:00 UTC (corresponding to 07:00 South African Standard Time - SAST) the RFI occupancy goes up, this may be related to when activities begin in the nearby towns and cities, and at times even on-site. At 10:00 UTC we see a drop in the RFI occupancy, similarly, at 14:00 UTC we see another drop. These two times correspond to lunch time and the end of the working day in South Africa respectively. We cannot conclusively say that the observed increase in RFI occupancy is caused by these human activities, however, a correlation clearly exists.\par\bigskip

We also found that the RFI probability at the following frequencies: 1018 MHz, 1031 MHz, 1041 MHz, 1090 MHz and 1103 MHz increases during the day time and drops at night time. These frequencies are confined within the DME band which is allocated to the aircraft communication system. Therefore, these findings suggest that the observed increase in RFI probability during the day is most probably due to the aircraft passing over a region of the site.\par\bigskip

Furthermore, there is a great deal of variation of RFI occupancy as a function of frequency at some frequency bands (e.g. 900 - 960 MHz) where we see $100\%$ RFI whereas at others (e.g. 1320-1500 MHz) the RFI occupancy is down to less than 10$\%$. We can see the three main frequency bands showing the highest probability of RFI in the MeerKAT site. Those are the Global System for Mobile Communication(GSM) (900 - 960 MHZ), aircraft transponders (1000 - 1200 MHz) and Global Positioning System(GPS) satellites (1482 - 1600 MHz $\&$ 1169 - 1280 MHz). From our analysis approximately 36.6$\%$ of the band at all the time, all the baseline is always flagged as RFI.\par\bigskip

In this paper, we are primarily interested in the RFI from known persistent sources such as GPS satellites, DMEs and GSM. Emission from these sources are fairly constant, predictable and regular. As a result, the variation in the probability of RFI from such sources is expected to be considerably small. For us to understand whether the observed fluctuations are statistically significant or are due to noise fluctuations, we computed the 68 percentile which corresponds to 1-sigma confidence interval for a Gaussian distribution. On the other hand, we suspect that the 95$\%$ confidence interval will include all sorts of outliers that may or may not be due to the radio signals. As a result, we found that the RFI variability for the GPS satellites, DMEs and GSM signals is more accurately captured by the 68$\%$ confidence limits.\par\bigskip

We further investigated the statistical consistency of the RFI probabilities. Figure \ref{time} shows the average RFI probability as a function of the time of the day, with the green region representing the $68\%$ confidence interval. We used two methods to calculate the average as explained in subsection \ref{analysis}. The blue line represents the CA, while the orange line represents the AoA. We observe a similar distribution of RFI from both methods.\par\bigskip

As mentioned earlier on this section, at 10:00 UTC  we observed a drop in the RFI occupancy. For this time of the day, we also found that the data is noisy as shown by the 68$\%$ confidence interval. To understand this noisiness, we looked at the distribution of the RFI probabilities at a noisier and quieter hour of the day.\par\bigskip

The huge variation of RFI probability in noisier hours is an indication of some form of an anomaly and this could be a result of several issues such as the correlator outputting zero visibilities. Indeed, we found that some of the observation had zero probabilities. The results imply that no RFI was detected on any baseline and at any frequency by the algorithm; something essentially impossible because of the permanent presence of RFI sources. This as mentioned above is an indication of a potential system problem, such as the correlator outputting zero visibilities. The SDP flagger does not detect any RFI when such events happen. Hence, we see a zero probability of observing RFI.\par\bigskip

\begin{figure*}
\begin{subfigure}[t]{0.4\hsize}
    \includegraphics[width=1.0\linewidth]{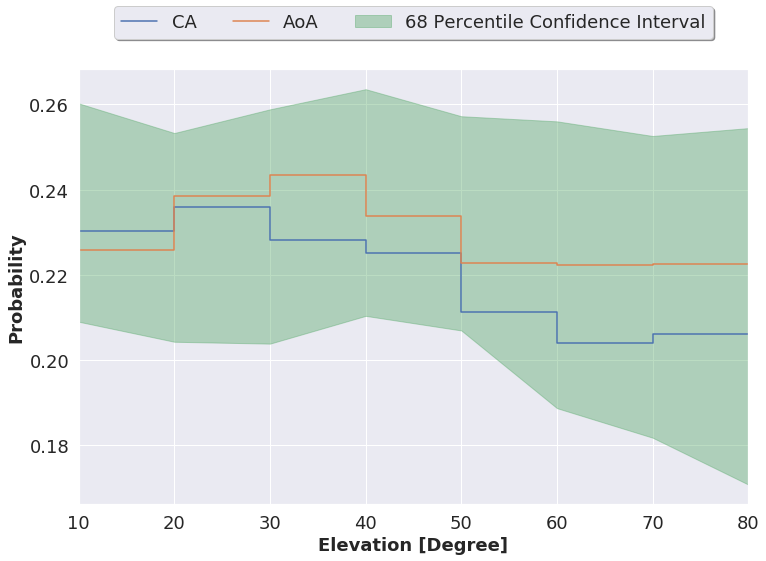}
    \caption{Elevation [Degrees]}
\end{subfigure}   
\begin{subfigure}[t]{0.4\hsize}
    \includegraphics[width=1.0\linewidth]{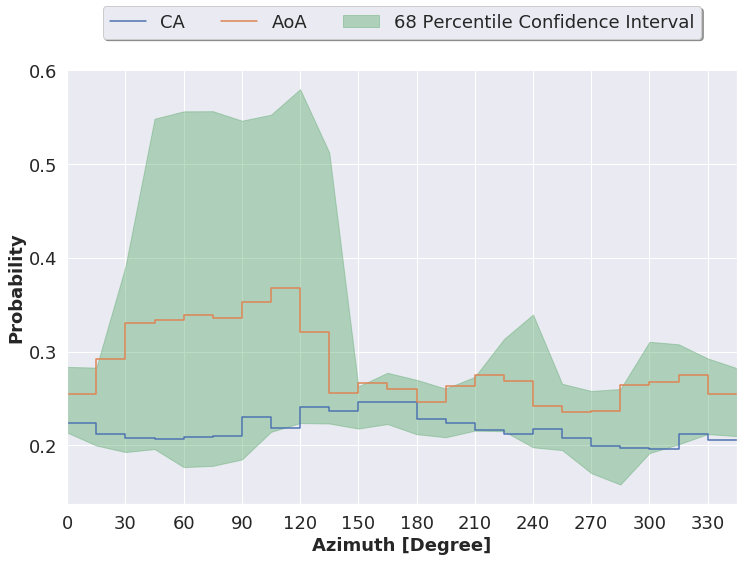}
    \caption{Azimuth [Degrees]}
\end{subfigure}
\caption[RFI occupancy for MeerKAT site as a function of telescope pointing with confidence intervals.]{RFI occupancy for MeerKAT site as a function of telescope pointing. The green region represents the 68$\%$ confidence interval. The blue and the orange lines represent the CA and AoA methods.  The confidence limits are wide for angles between $30^\circ$ and $140^\circ$ on the azimuth plot.}
\label{pointings_with_error}
\end{figure*}

We performed a similar analysis on the frequency axis, however, we decided to split the frequency spectrum into a known corrupted band and a clean band. The corrupted band is defined as the range of frequencies in which the major known RFI sources (GSM, DME and GPS satellites) emit, whereas the clean band is the less corrupted part of the spectrum. We sub-divided the clean band into lower and upper frequencies which are, between 980 MHz - 1070 MHz and 1310 MHz - 1500 MHz respectively. This was done by inspecting the RFI contribution as a function of  frequency, Fig.~\ref{fig:time_freq}. Figure \ref{fig:freq_95_68} shows the RFI averages with the blue and the orange line being computed from the CA and AoA method respectively. Meanwhile the green region represent the  68$\%$ confidence interval. We noticed a small variation in the RFI occupancy in the corrupted band as shown by the 68$\%$ confidence interval limits which are tightly constrained around the mean. However, as for the lower and the upper clean bands we do find frequencies (e.g. 1030 MHz, 1040 MHz, 1381 MHz, 1390 MHz and 1492 MHz) in which the RFI occupancy is greater than 10$\%$, these are depicted by spikes in those regions. We observe a relatively high variation in RFI occupancy at these particular frequencies when looking at the 68$\%$ confidence interval.\par\bigskip

We, therefore, looked at the distribution of probabilities of some of the clean band frequencies. We expected the distribution of the RFI probabilities in the clean band to be close to zero, as there should not be any contamination. However, we see a long tail distribution towards higher values of RFI probability. This long tail is a result of rare events that are appearing much more frequently than we expected. For example, the 1380 MHz L3 GPS band which is used for detecting nuclear activity on Earth seems to have been more active. The two frequencies shown are confined within the GPS L3 band.\par\bigskip
\begin{figure}
\centering
\includegraphics[width=9cm]{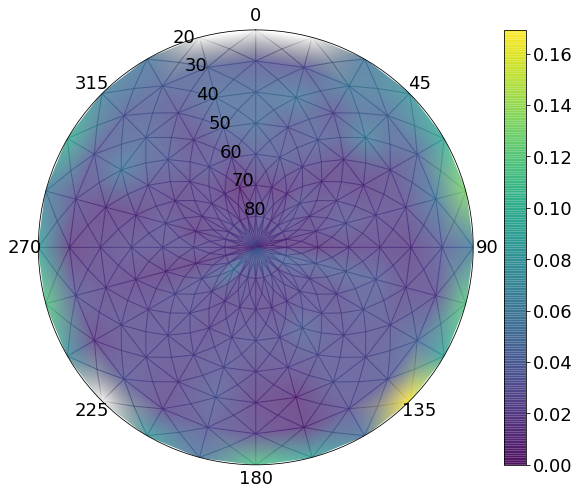}
\caption{RFI occupancy as a function of the telescope pointing direction for the  clean band. We can notice a hot spot at low elevation and azimuth of 135$^\circ$ which is pointing towards nearby towns.}
\label{fig:clean_band_direc}
\end{figure}

\begin{figure}
    \centering
     \includegraphics[width=9cm]{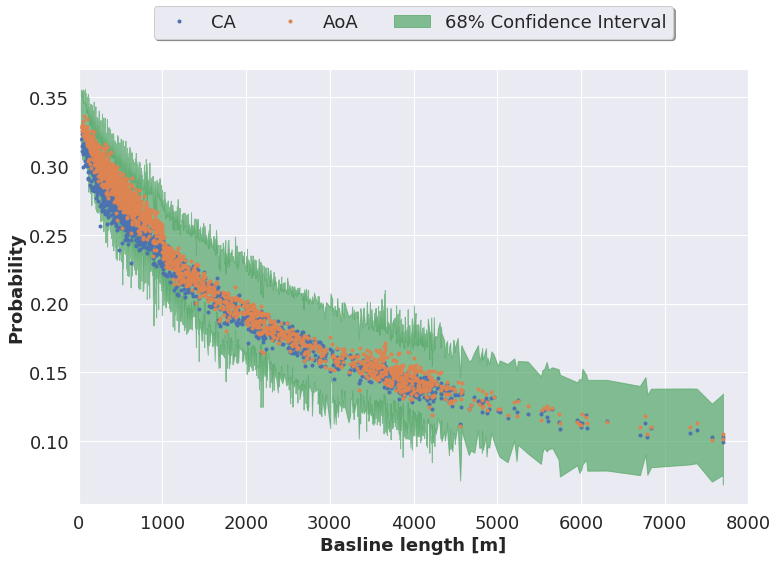}  
    \caption[Probability of RFI for the MeerKAT telescope as a function of Baseline length.]{RFI occupancy for the MeerKAT telescope as a function of Baseline length (m). The blue and the orange dots are the mean RFI probability from the CA and AoA averages respectively, while the green region represents the 68$\%$ confidence interval. The decrease of the RFI probability with  increase of baseline length is due to moving RFI sources with respect to the static sky which causes the phase of the RFI to oscillate rapidly on long baselines compared to short baselines which then tend to progressively average out on longer baselines when the visibilities are averaged over typical timescales (0.5-8s for MeerKAT).}
    \label{fig:baseline}
\end{figure}

Furthermore, we looked at how the RFI occupancy changes as the telescope points at various directions in the sky. The amount of RFI it measures is expected to change depending on the number of radio frequency transmitters that is in the field of view. It is anticipated that RFI due to terrestrial sources should be more dominant at low elevation. Figure \ref{pointings_with_error} was used to examine this possibility. We noticed that between $20^\circ$ and $50^\circ$ Elevation the RFI probability is the highest and it gradually drops as we go to higher elevations on both the CA and AoA methods. This results can explain that indeed at low Elevation we do see more RFI as compared to high elevations.\par\bigskip

Likewise, we computed the  68$\%$ confidence interval for the elevation axis and the azimuth axis, Fig \ref{pointings_with_error}. We found that the  68$\%$ confidence interval limits on the elevation are tightly constrained around the mean, hence a small variation in RFI probability is observed. As for the azimuth plot we found that some of the directions ($30^\circ$ and $140^\circ$) are too noisy. In order for us to understand the observed large variations we took a slice at a specific direction to look at the distribution of the RFI probabilities.\par\bigskip

Looking at the number of counts for both noisier and quieter azimuth we noticed that the noisier angle has less count. It is worth noting that the count on outliers is comparable for both angles. Thus, this is indicative of the lack of data in those regions.\par\bigskip

The polar plot in Figure \ref{fig:clean_band_direc} shows how much RFI is generated in the clean band as a function of Azimuth (radial direction) and Elevation (theta direction). MeerKAT uses a lower limit in the elevation and we have chosen 20$^\circ$ as our lower limit (since most observations had this common lower limit). The white empty areas (Azimuth: $225^\circ$ - $240^\circ$ and $345^\circ$ - $360^\circ$) are indicative of lack of data for these angles in our analysis. The colour scale ranging from purple through blue to yellow represents the probability of RFI occupancy, with yellow denoting the highest probability while purple is representing the lowest probability of RFI.\par\bigskip

We noticed a hot-spot (maximum RFI occupancy) at lower elevations and azimuth angle of 135$^\circ$ that coincidentally points towards the town Beaufort West which is nearby the MeerKAT site. In addition, the RFI occupancy is quite moderate across the azimuth angles at lower elevations. Looking at higher elevations (elevations > 40$^\circ$) the average RFI occupancy is about 2$\%$. We require further investigations to confirm the sources of these RFI.\par\bigskip

Finally, we investigated the probability of RFI as a function of baseline length, Fig \ref{fig:baseline}. The blue and orange dots are the average RFI probabilities from the two different averaging methods, CA and AoA respectively. Meanwhile the green region represents the 68$\%$ confidence interval. We notice that the RFI probability decreases as a function of baseline length from both methods.\par\bigskip

To explain the observed decrease in RFI probability as a function of the baseline length, consider the complex visibility:
\begin{eqnarray}\label{van}
V(u,v) = \int \int I(l,m)  e^{-2\pi i(ul + vm )} dldm
\end{eqnarray}
of a single source which is produced by the multiplying the sky ($I(l,m)$) with the fringe pattern produced by the baseline integrated over solid angle. The angular distance between two consecutive peaks of the fringe pattern is defined as the fringe spacing. The fringe spacing is dependent on the separation between the antennas, with a short baseline giving a large fringe spacing while the long baseline gives smaller fringe spacing.\par\bigskip

For RFI sources that are moving with respect to the static sky, the phase of these RFI sources wraps rapidly on long baselines compared to the phase for short baselines. Therefore, when a correlation is carried out on long baselines the RFI amplitude averages incoherently and is reduced. On the contrary, the short-baselines tend to add coherently, hence as a result, when the correlation is carried out the RFI amplitude is reduced less than on longer baselines,~\cite{offringa2013lofar}.\par\bigskip

Overall, from these preliminary findings we can say that the RFI environment is dynamic. The clean band is supposed to be as clean as possible from RFI, but collectively, our results show evidence of activities that are happening which are worth investigating in the future.

\section{Conclusions}

Radio astronomers typically flag RFI and outliers from their data without caring much about the origin and source of the contamination.  On the other hand, radio observatories are very interested in accurately characterising and understanding the RFI environment around the observatory to meet the goal of ensuring the best quality of data possible from the telescope.\par\bigskip

We have presented a framework that provides a multi-dimensional statistical view of the RFI environment of an observatory using the data from the telescope array itself with an automated flagger of the RFI. This approach can be applied to any archival data from observatories to understand the evolution and nature of the RFI.\par\bigskip
 
Using around 1500 hours of MeerKAT telescope array data as a demonstration we produce the RFI and outlier occupation probabilities over several months as a function of time of the day \textbf{(T)}, frequency channels \textbf{(F)}, baseline length \textbf{(B)}, elevation \textbf{(El)} and azimuth \textbf{(Az)}. Our framework  presented here can be adapted to any radio telescope. Beyond its use for alerting to new sources of RFI and understanding trends in the RFI environment, our results can provide useful prior probabilities for RFI flagging. For example, an observer interested in specific lines (e.g. the 21cm line) for which there is danger of confusion with RFI, can use our multi-dimensional array to compute the prior probability that an observed spike is astronomical or RFI and hence reduce both contamination and missed signal. 

\section*{Acknowledgments}
 
We thank Christopher Finlay, Dr Tom Mauch, the SARAO data science team,  and the MeerKAT RFI working group for discussions and support during this project. The MeerKAT telescope is operated by the South African Radio Astronomy Observatory, which is a facility of the National Research Foundation, an agency of the Department of Science and Innovation. This research has been conducted using resources provided by the United Kingdom Science and Technology Facilities Council (UK STFC) through the Newton Fund and the South African Radio Astronomy Observatory.


\bibliographystyle{mnras}
\bibliography{ref}

\begin{thebibliography}{}
\makeatletter
\relax
\def\mn@urlcharsother{\let\do\@makeother \do\$\do\&\do\#\do\^\do\_\do\%\do\~}
\def\mn@doi{\begingroup\mn@urlcharsother \@ifnextchar [ {\mn@doi@}
  {\mn@doi@[]}}
\def\mn@doi@[#1]#2{\def\@tempa{#1}\ifx\@tempa\@empty \href
  {http://dx.doi.org/#2} {doi:#2}\else \href {http://dx.doi.org/#2} {#1}\fi
  \endgroup}
\def\mn@eprint#1#2{\mn@eprint@#1:#2::\@nil}
\def\mn@eprint@arXiv#1{\href {http://arxiv.org/abs/#1} {{\tt arXiv:#1}}}
\def\mn@eprint@dblp#1{\href {http://dblp.uni-trier.de/rec/bibtex/#1.xml}
  {dblp:#1}}
\def\mn@eprint@#1:#2:#3:#4\@nil{\def\@tempa {#1}\def\@tempb {#2}\def\@tempc
  {#3}\ifx \@tempc \@empty \let \@tempc \@tempb \let \@tempb \@tempa \fi \ifx
  \@tempb \@empty \def\@tempb {arXiv}\fi \@ifundefined
  {mn@eprint@\@tempb}{\@tempb:\@tempc}{\expandafter \expandafter \csname
  mn@eprint@\@tempb\endcsname \expandafter{\@tempc}}}

\bibitem[\protect\citeauthoryear{Asad et~al.,}{Asad
  et~al.}{2019}]{asad2019primary}
Asad K.,  et~al., 2019, arXiv preprint arXiv:1904.07155

\bibitem[\protect\citeauthoryear{Camilo et~al.,}{Camilo
  et~al.}{2018}]{camilo2018revival}
Camilo F.,  et~al., 2018, The Astrophysical Journal, 856, 180

\bibitem[\protect\citeauthoryear{Carter, Ant{\'o}n, Dagnino  \&
  Williams}{Carter et~al.}{2001}]{carter2001evolving}
Carter R.~A.,  Ant{\'o}n A.~I.,  Dagnino A.,   Williams L.,  2001, in
  Proceedings Fifth IEEE International Symposium on Requirements Engineering.
  pp 94--101

\bibitem[\protect\citeauthoryear{Fridman \& Baan}{Fridman \&
  Baan}{2001}]{fridman2001rfi}
Fridman P.,  Baan W.,  2001, Astronomy \& Astrophysics, 378, 327

\bibitem[\protect\citeauthoryear{Mauch et~al.,}{Mauch
  et~al.}{2020}]{mauch20201}
Mauch T.,  et~al., 2020, The Astrophysical Journal, 888, 61

\bibitem[\protect\citeauthoryear{Offringa, De~Bruyn, Biehl, Zaroubi, Bernardi
  \& Pandey}{Offringa et~al.}{2010}]{Offringa2010post}
Offringa A.,  De~Bruyn A.,  Biehl M.,  Zaroubi S.,  Bernardi G.,   Pandey V.,
  2010, Monthly Notices of the Royal Astronomical Society, 405, 155

\bibitem[\protect\citeauthoryear{Offringa et~al.,}{Offringa
  et~al.}{2013}]{offringa2013lofar}
Offringa A.,  et~al., 2013, Astronomy \& astrophysics, 549, A11

\makeatother
\end{thebibliography}

\label{lastpage}
\end{document}